\documentclass[twocolumn]{aa}
\usepackage{psfig}
\usepackage{natbib}
\usepackage{amssymb}
\bibpunct{(}{)}{;}{a}{}{,}
\usepackage{ltxtable}
\usepackage[latin1]{inputenc}
\newcommand{\dpart}[2]{\frac{\partial #1}{\partial #2}}

\newcommand{\etal}{et al.}

\def\pmb#1{\setbox0=\hbox{#1}%
  \kern-.025em\copy0\kern-\wd0
  \kern.05em\copy0\kern-\wd0
  \kern-.025em\raise.0433em\box0 }

\begin{document}
   \title{The dynamical influence of cooling in the envelope of
   prestellar and protostellar cores.}

   \author{P. Lesaffre \inst{1,2,3}         
        \and
     A. Belloche \inst{1,4}
	\and
	J.-P. Chièze \inst{1}
\and	
	P. André \inst{1}
	  }

   \offprints{lesaffre@ast.cam.ac.uk}

   \institute{CEA/DAPNIA/SAp Orme des Merisiers, F-91191
   Gif-sur-Yvette Cedex 
 \and Institute of Astronomy, Madingley Road, Cambridge 
CB3~0HA, UK 
 \and University of Oxford, Department of Astrophysics,
 Oxford OX1~3RH, UK
 \and Max-Planck-Institut f\"ur Radioastronomie, Auf
 dem H\"ugel 69, 53121 Bonn, Germany }

 \date{Received September 15,
 1996; Accepted March 16, 1997}

   \abstract{ 

We compute numerical simulations of spherical collapse triggered by a
slow increase in external pressure. We compare isothermal models to
models including cooling with a simple but self-consistent treatment
of the coupling between gas, grains and radiation field temperatures.
The hydrostatic equilibrium appears to hold past the marginally stable
state, until the collapse proceeds.  The last hydrostatic state before
collapse has a lower central gas temperature in the centre due to the enhanced
coupling between gas, grains and radiation field. This results in
slightly lower pressure gradients in the bulk of the envelope which is
hence slightly more extended than in the isothermal case.  Due to the
sensitivity of the collapse on these initial conditions, protostellar
infall velocities in the envelope turn out to be much slower in the case
with cooling.

Our models also compute the radiative transfer and a rather large
chemical network coupled to gas dynamics.  However, we note that the
steady-state chemisorption of CO is sufficient to provide an accurate
cooling function of the gas. This justifies the use of post-processing
techniques to account for the abundance of observed molecules.

Existing observations of infall signatures put very stringent
constraints on the kinematics and temperature profile of the class 0
protostar IRAM~04191+1522. We show that isothermal models fail to account
for the innermost slow infall motions observed, even with the most
hydrostatic initial conditions. In contrast, models with cooling
reproduce the general shape of the temperature profile inferred from
observations and are in much better agreement with the infall
signatures in the inner 3000~AU.

}

   \maketitle

\section{Introduction}

  We revisit here the numerical work of \citet{L69}, \citet{M98} and
  \citet{MI00} to probe the influence of cooling and chemistry on the
  collapse of protostellar and prestellar dense cores.  We compare
  our results to observations of a Class 0 protostar, i.e., a
  collapsing protostellar core so young that it still retains a detailed
  picture of its initial conditions \citep{A99}.

 Isothermal studies of spherical collapse have focused on self-similar
 solutions \citep{S77,L69,B85,B88}, initial conditions \citep{FC93}
 and magnetic fields \citep{BM94, Li98}.

  In contrast, simulations of hydrostatic clouds
  \citep{D80,BD84,CP87,NW99} have paid extreme attention to the energy
  budget, using accurate models of radiative transfer coupled with
  complicated chemical networks designed to account for the abundances
  of the main cooling agents of the interstellar medium. Using more
  simple thermal models, \cite{G02} have also investigated hydrostatic
  equilibria near the marginally stable state.

  In the present work, we use a similar approach to \citet{MI00}. We
  use a simplified treatment of the radiation transfer, which is found
  to give very similar results. This allows us to spend the CPU time
  on a more detailed chemical network. We are hence able to model the
  molecular cooling with a degree of accuracy close to the
  hydrostatic models. Also, we use a moving grid algorithm
  which describes the accretion shock with a much better resolution.

  The aim of this paper is to show how the use of a more accurate
  model for the gas and dust microphysics can influence the dynamics
  of the protostellar collapse.  We briefly present isothermal models
  (Section 2) before comparing them to the results of computations
  including radiation transfer and using a full chemical network
  (Section 3).  We emphasise the importance of the coupling model
  between gas and radiation through the grains which allows us to
  compute a realistic profile for the temperature. This profile is
  responsible for slightly lower pressure gradients which make the
  collapse milder than in the isothermal case.  We investigate the
  influence of a few parameters and in particular show that the 
  out-of-equilibrium chemistry has little influence on the 
  cooling of the gas.  We finally compare our models to observations
  of the young low mass class 0 protostar IRAM~04191 (Section 4). We
  find that isothermal models fail to reproduce the slow observed
  infall velocities. On the other hand, models with cooling are within
  the observational uncertainties inside a radius of around 3000~AU.
  We discuss and sum up our results in sections 5 and 6.

\section{Isothermal models}

\subsection{Initial conditions}
	
We start our collapse calculations from a stable hydrostatic state. We
then trigger the gravitational instability by a slow exponential
increase of the external pressure $p_{\rm ext}$. When the critical
pressure is reached, the collapse evolves on a few free-fall time
scales, until the birth of the protostar. A jump in the mass-to-radius
ratio at the centre characterises this particular instant, which we
take as the origin of times $t=0$ \citep[as in][]{FC93}. In the
present work, we refer to times $t<0$ as the prestellar phase and
$t>0$ as the protostellar phase, in agreement with the terminology
used by most observers.  Three parameters specify a simulation~: the
total mass $M$, the temperature $T$, and the time scale for the
pressure increase $t_p=$d$t/$d$\ln(p_{\rm ext})$. We use
$\mu=2.33$~amu for the mean molecular weight .

  Because the equations can be put in an adimensional form
  \citep{FC93} only the parameter $t_p$ is relevant. We chose to fix
  $M=1.7~$M$_\odot$ and $T=10$~K as good estimates for 
  the object we plan to compare to (see section
  \ref{compiso}). The free-fall time-scale of the marginal
  Bonnor-Ebert sphere with these parameters is $t_{\rm
  ff}=1.3 \times 10^5$~yr.

\subsection{Method}
  The use of a Lagrangian mesh allows us to easily treat the huge
  dynamical range in densities and scales involved in the collapse
  without introducing advection errors. However, it is not able to
  evolve through the central singularity. We therefore use a moving
  grid algorithm that keeps the mesh on a logarithmic spacing in
  radius with a fixed radius for the central zone when it gets below
  $10^{-8}$~pc. To check this method, we compared our prestellar
  results with the output of a Lagrangian code. The comparison is
  worse for the maximum infall velocity which is usually about 10\%
  smaller than in the Lagrangian computation. The collapse 
  outside of the point of maximum infall velocity
  is reproduced to an accuracy better than 1\%. Our
  protostellar results agree with \cite{FC93} to an accuracy better
  than 5\% when we use their initial conditions, i.e.  a 10\%
  enhancement of pressure and density compared to the marginally
  critical Bonnor-Ebert sphere. We note that they do not enhance the
  pressure in their outer buffer region. Our simulations have also
  been compared to the slow pressure increase case of \cite{H03}
  with a satisfactory agreement.
\label{method}

\subsection{Supersonic mass fraction}
\begin{figure}[h]
\centerline{
\psfig{file=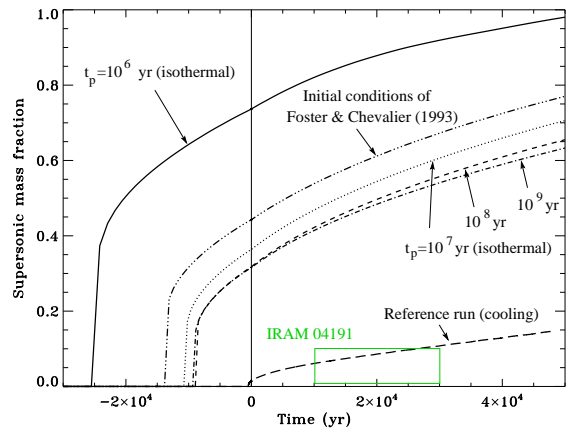,width=9cm,angle=-90}
} 

 \caption{Evolution of the supersonic mass fraction $f_s$ over time
 for different initial conditions of an isothermal collapse. Also
 shown is the mass fraction with motions greater than the surface
 sound speed for our reference run with cooling. The thin solid line
 is the mass over radius ratio at centre in non-dimensional units
 \citep[$m/\xi$ as defined by][]{FC93} for the isothermal model with
 $t_p=10^6$~yr: it is almost vertical and hence makes a precise
 indicator for the instant $t=0$. The observational constraints set by
 \citet{B02} for the class 0 protostar IRAM~04191 are shown as a
 rectangle.  }
\label{x2}
\end{figure}

Figure \ref{x2} shows the evolution of $f_s$, the fraction of mass
with supersonic infall motion, during our isothermal runs and during
the simulation with the same initial conditions as \citet{FC93} (see section
\ref{method}).
$f_s$ depends critically on the initial conditions and on the
parameter $t_p$: the faster the increase in pressure, the more
dynamical the collapse.  This makes it a good test for the accuracy of
the code and we reproduce the $f_s(t=0)=0.44$ result of \citet{FC93}.

\section{Non isothermal models with chemistry}

\subsection{Numerical and physical inputs}
\subsubsection{Method}
  We use an improved version of a code designed to compute shocks in
  the interstellar medium \citep{Les04}. This code makes use of a
  moving grid algorithm which resolves at best temperature and
  chemical gradients for a fixed number of 100 zones. It also solves
  for multifluid magneto-hydrodynamical equations fully coupled to a
  chemical network. In the present version of the code, we set the
  magnetic field to zero, wrote the equations in spherical coordinates
  and added the radiative transfer equations and the energy budget for
  the grains.

  In the following we mention only a few additional features that we
  included in the code~:
\begin{itemize}
\item adsorption and desorption chemistry onto and from grains.
\item transfer of the continuum radiation and approximate treatment of
line cooling.
\item energy budget of the grains which transfer energy
between gas and radiation.
\end{itemize}

  Further details of the method can be found in \citet{Les02} and
\citet{A02}.

\subsubsection{Chemistry}
  The chemical network used in \citet{Les04} comprised 120 reactions
  for 32 species.  This network was designed to compute the abundances
  of the main molecular and atomic cooling agents of the interstellar
  medium.  We added 4 species adsorbed onto grains~: CO, H$_2$O,
  CH$_4$ and O$_2$.  We describe their rate of adsorption onto grains
  using a model similar to \citet{NW99}. For the sticking coefficient,
  we use expression (4) of \citet{HS71} calibrated by \citet{BZ91} and
  \citet{M98}. We model the desorption due to cosmic-rays using the
  hot spot model of \citet{L85}.  The formation of  H$_2$ molecules
  is the only reaction on grain surfaces included. Reactions of
  photo-dissociation, photo-ionisation as well as thermal sputtering
  of the grains are not included. Our treatment is thus accurate only
  up to grain temperatures of around 100~K.  In the present work, we
  refer to the core as the innermost few AU where grains have such
  warm temperatures and are optically thick to radiation. We refer to
  the rest of the cloud as the envelope.

\subsubsection{Radiation field}

  The transfer for the continuum radiation is done according to
  \cite{DF99} as in \cite{A02} with the Eddington factor obtained from
  the minimisation of the entropy of the photon gas. The Planck mean
  opacities  of the grains are from \citet{AS86} \citep[ie:
  similar to ][]{DL84}. The outer boundary condition is a fixed
  temperature $T_r=2.7$~K and the inner one is a zero flux. Note that
  by $T_r$ we mean the {\it bolometric} temperature of the photon gas
  such that the radiation energy density is $E=aT_r^4$ where $a$ is
  the radiation density constant.

  The line transfer is assumed to be optically thin for the atoms and
  H$_2$. This is not a good approximation deep in the core but this
  cooling turns out to be negligible there.  The other sources of molecular
  coolings are from \citet{HM79} (for OH) and \citet{NK93} (CO and
  H$_2$O). For the optical depth parameter $\tilde{N}(\rm M)$ of
  molecule M \citep[see][]{NK93} we use a combination of the
  column density of these molecules as well as the velocity gradients
  so that radiation takes the easiest way to escape~:
\begin{equation}
  \tilde{N}(\rm M)=n(\rm M)/v'\mbox{,}
\end{equation} 
with
\begin{equation}
v'=\sqrt{(\dpart{v}{r})^2+(\frac{c_s}{d})^2}
\end{equation}
and
\begin{equation}
d=R-r+A_v(R)/[1.5 \times 10^{-21}{\rm cm}^2 \times n_H(R)] \mbox{.}
\end{equation}
  $n(\rm M)$ is the local density of molecule M, $v$ is the velocity
  of the gas and $c_s$ its sound speed, $r$ is the distance to the
  centre, $R$ is the outer radius, $n_H$ is the density of H nuclei
  and $A_v(R)$ is the extinction in the visible at the outer edge of
  the cloud ($A_v(R)=1$ or 2 in this study).

  As discussed by \citet{HM79}, the escape probabilities of the
  photons should be modulated by the probability for each line that an
  escaping photon be absorbed onto a grain. However, their treatment
  is only valid as long as the grains are optically thin. This is not
  valid anymore in the context of star formation after the first
  adiabatic core has formed.  The correct treatment would involve a
  multi-group treatment of the radiation field with a very fine grid
  in wavelength. This is still way out of today's computational
  capabilities. However, we tested two extreme hypotheses~: either the
  line radiation is all absorbed by grains or the grains are
  transparent to the line radiation.  We found that the collapse in
  the envelope proceeds in a very similar manner for both
  hypotheses. In the following, the results are presented for grains
  transparent to the line radiation.

  Finally, the absorption of the UV field that enters only the
  expression for the heating due to the photoelectric effect is
  treated implicitly through expressions (\ref{photel}) and (\ref{Av})
  below.

\subsubsection{Grain temperature}

 The heating and cooling processes included for the grains are~:
\begin{itemize}

 \item absorption and emission of radiation by the grains
 (including the line cooling or not). The net energy rate gained
by the grains is~:
\begin{equation}
\Gamma_{\rm \tiny radiation}=ac \mu \kappa(T_d) n_H(T_r^4-T_d^4)
+\beta \Gamma_{\rm \tiny lines}
\end{equation}
 where $T_r$, $T_d$ are the radiation and dust temperatures, $\mu$ is
 the mean weight per hydrogen nucleus, $\kappa(T_d)$ is the Planck
 opacity of the grains \citep[from][]{AS86} and $c$ is the speed of
 light. $\Gamma_{\rm \tiny lines}$ is the rate of energy lost by the
 gas (hence gained by the grains) via emission through
 lines. $\beta=1$ when lines are assumed to be absorbed onto
 grains. $\beta=0$ when grains are transparent to the lines.\\

 \item collisional exchanges with the gas~:
\begin{equation}
\label{coll}
\Gamma_{\rm \tiny collisional}=\alpha n_H^2 T_g^{\frac12}(T_g-T_d)
\end{equation}
with $\alpha=3.5~10^{-34}~$ in cgs units \citep[see][ page 731]{B87}
and $T_g$ is the gas temperature.\\

 \item heat released by H$_2$ formation (2/3 or 1/3 of H$_2$'s binding
 energy, depending if the line cooling is absorbed by the grains or
 not)~:
\begin{equation}
   \Gamma_{{\rm H}_2}=R_f Q (\beta+1)/3
\end{equation}
 where $R_f$ is the rate of formation of H$_2$ molecules on grains 
and $Q=4.48$~eV is the binding energy per H$_2$ molecule.\\

 \item photoelectric effect on the grains is from \citet{B87}, page 731~:
\begin{equation}
\Gamma_{\rm \tiny photoelectric}=4 \times 10^{26} G_0 \exp(-2.5A_v)n_H
\label{photel}
\end{equation}
where $ G_0 $ measures the intensity of the ambient UV radiation field 
compared to the interstellar mean UV-field. $ G_0 $ is set to 1 and
$A_v$ is computed from~:
\begin{equation}
A_v(r)=A_v(R)+1.5 \times 10^{-21}{\rm cm}^2 \int_r^R n_H {\rm d}r
\label{Av}
\end{equation}
\end{itemize}

  The temperature of the grains is obtained by  setting their 
energy budget to zero:
\begin{equation}
 \Gamma_{\rm \tiny radiation}+\Gamma_{\rm \tiny collisional}
 +\Gamma_{{\rm H}_2}+\Gamma_{\rm \tiny photoelectric}=0 \mbox{.}
\end{equation}

Inside the collapsing cloud, the central extinction $A_v(0)$ rapidly
increases and shields the photoelectric effect.
  If we further neglect the absorption of the lines by grains and the
  heat released by H$_2$ formation, the energy budget of the grains
  simply reads~:
\begin{equation}
ac \mu \kappa(T_d) n_H(T_r^4-T_d^4)-\alpha n_H^2 T_g^{\frac12}(T_d-T_g)=0
\label{budget}
\end{equation}
 The grain temperature hence lies
  between the gas and the radiation temperature. 

Equation (\ref{budget}) becomes
\begin{equation}
  n_0(T_r-T_d)=n_H(T_d-T_g)
\label{sbudget}
\end{equation}
  if we set
\begin{equation}
n_0=ac\mu\kappa\alpha^{-1}(T_d^3+3T_d^2T_r+3T_dT_r^2+T_r^3)T_g^{-\frac12}
\mbox{.}
\end{equation}
  For densities much greater than $n_0$ the grain temperature is well
  coupled to the gas; for densities much lower than $n_0$ the grain
  temperature is close to the radiation temperature.  However, the
  temperature dependence of $n_0$ often leads to a large range of
  densities for which the grain temperature is close neither to $T_g$
  nor to $T_r$ (especially for low $T_r$). For $\mu=2.33$~amu,
  $\kappa=10^{-2}~$cm$^2.$g$^{-1}$ (at $T_d=10$~K) and
  $T_d=T_g=T_r=10$~K we get
  $n_0=6.3~10^7~$cm$^{-3}$. Many authors \citep[see][ for example]{C96,G01,G02}
  neglect $\Gamma_{\rm \tiny collisional}$ in the grains energy budget
  when they estimate the grain temperature. Then they compare the
  radiative cooling of the gas to the collisional exchanges between
  gas and grains and find a critical density of $10^5$~cm$^{-3}$ for
  the coupling between gas and grain temperature, much lower than
  $n_0$.  It is hence important to consider at least all terms in
  equation (\ref{budget}) rather than neglecting the gas-dust
  collisional exchanges in the grains budget.

  In effect, the dust grains serve as a mediator for the energy
  transfer between gas and radiation. Indeed, plugging (\ref{sbudget})
  into (\ref{coll}) yields the effective heating for the gas through
  gas-dust collisions expressed as a coupling between radiation and
  gas temperatures:
\begin{equation}
\label{effective}
\Gamma_{\rm \tiny dust \rightarrow gas}=\alpha n_H^2 (\frac{n_H}{n_0}+1)^{-1}
T_g^{\frac12}(T_r-T_g) \mbox{ .}
\end{equation}

\subsection{Results}

  As in the isothermal case, we start from a hydrostatic state and
  trigger the collapse with a slow increase in external
  pressure. Since we now compute the temperature of the gas
  self-consistently, the parameter $T$ of the isothermal models is no
  longer relevant and is replaced by the outer boundary conditions for
  the radiation fields. Our reference run uses $T_r=2.7~$K, $ G_0 =1$,
  $A_v(R)=1$, a total mass of 1.7~M$_\odot$ and a pressure increase
  time scale $t_p=10^7$~yr.  Another parameter is the primary
  ionisation rate by cosmic-rays $\zeta$ which we fixed to
  $5 \times 10^{-17}$~s$^{-1}$. This parameter is important in the chemistry and
  for the heating of the gas through cosmic-rays \citep{SS69}.  In the
  following, we describe the time evolution of this reference model,
  compare this model to the isothermal models and investigate the
  influence of several parameters.

\subsubsection{Description of the reference model}
\label{reference}

\begin{figure*}[h]
\begin{center}
\begin{tabular}{cc}
\psfig{file=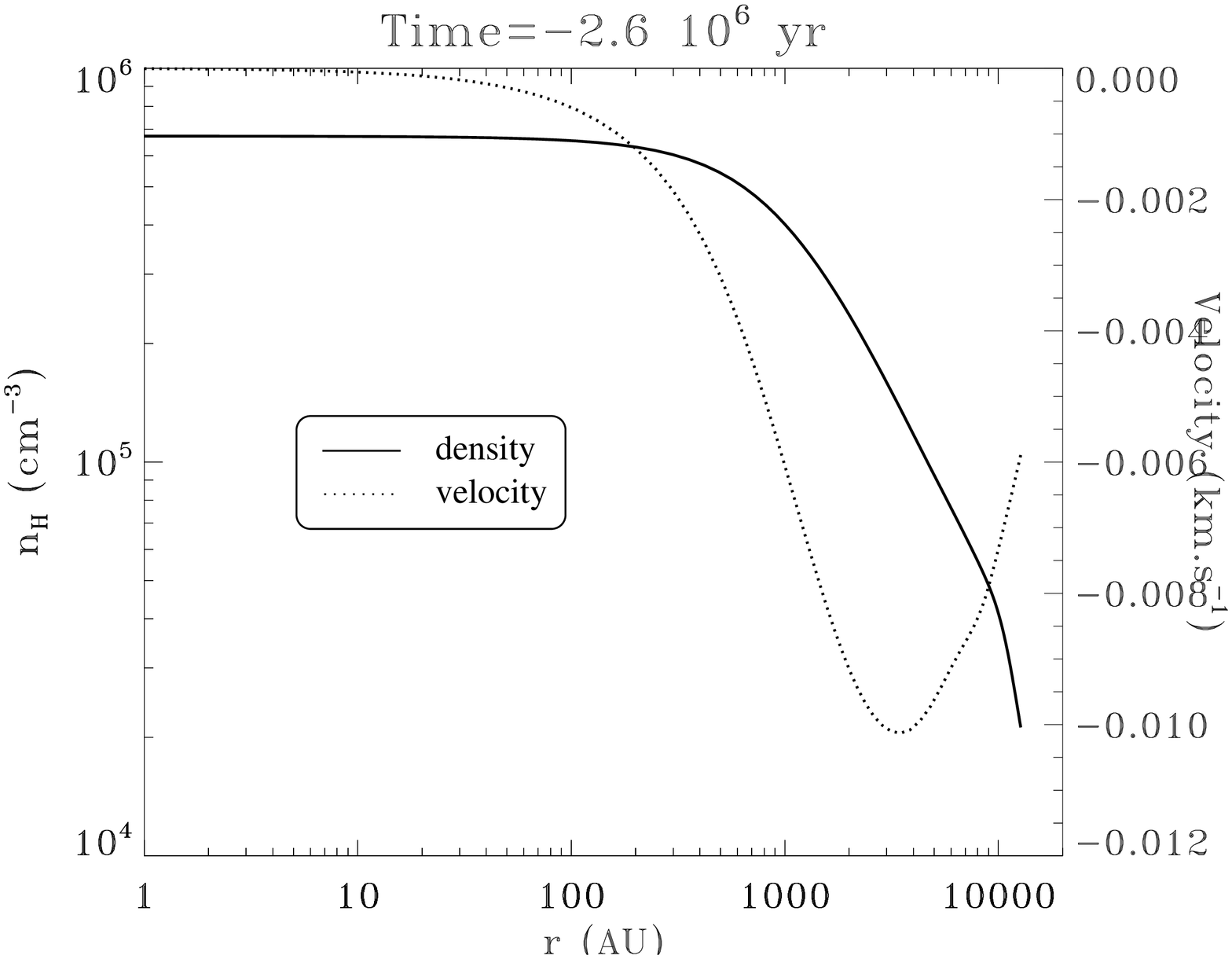,width=8cm,angle=-90}&
\psfig{file=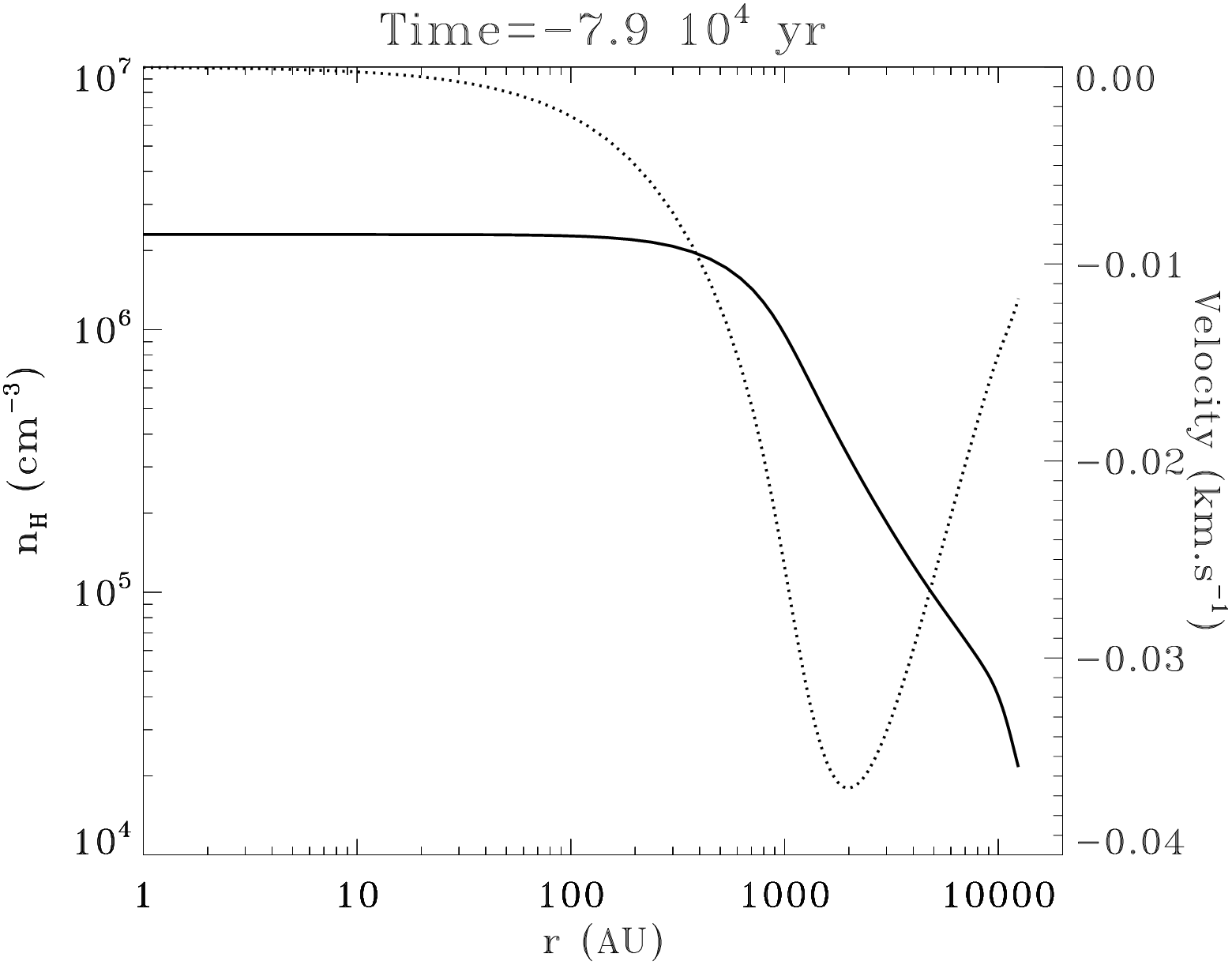,width=8cm,angle=90}\\
\psfig{file=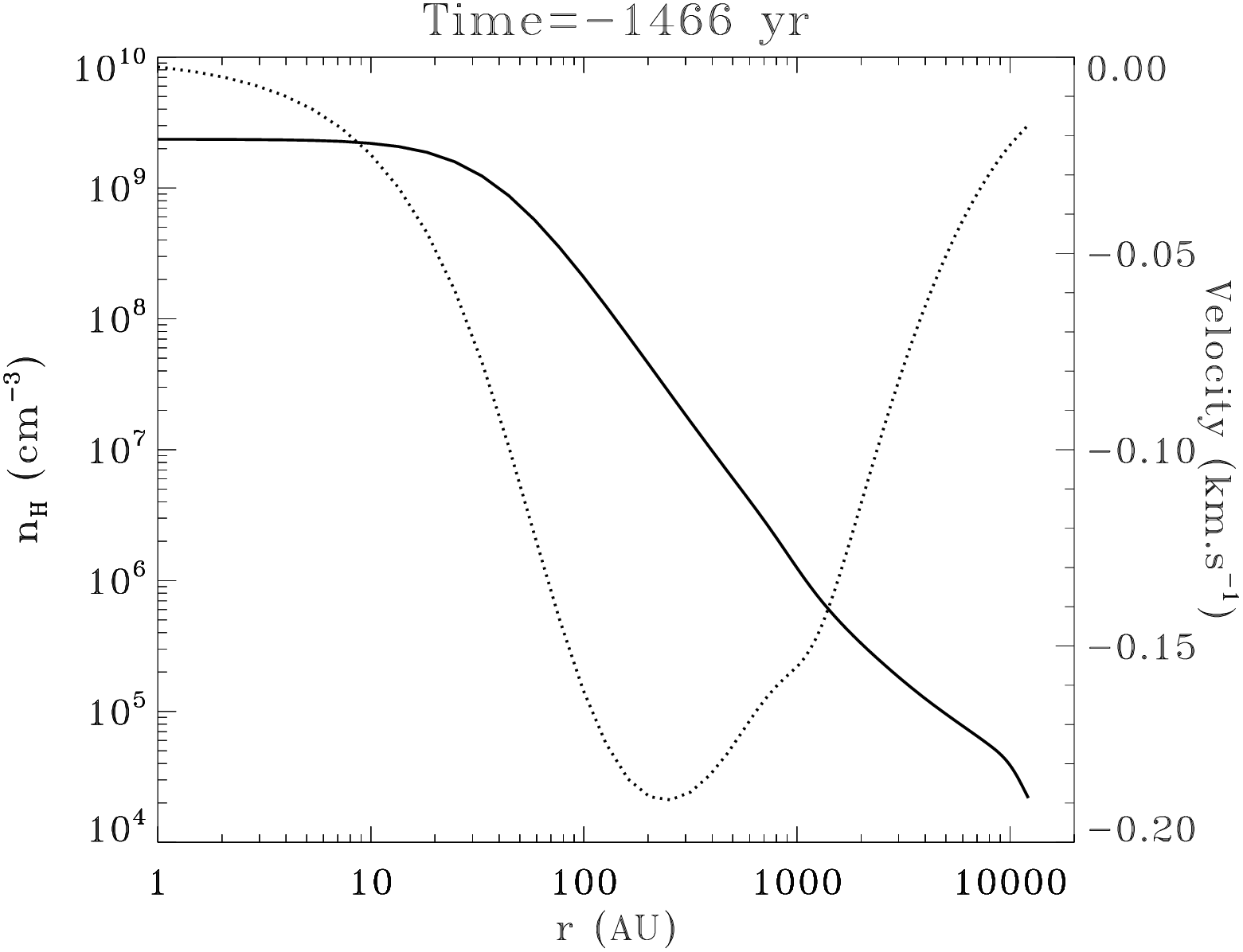,width=8cm,angle=90}&
\psfig{file=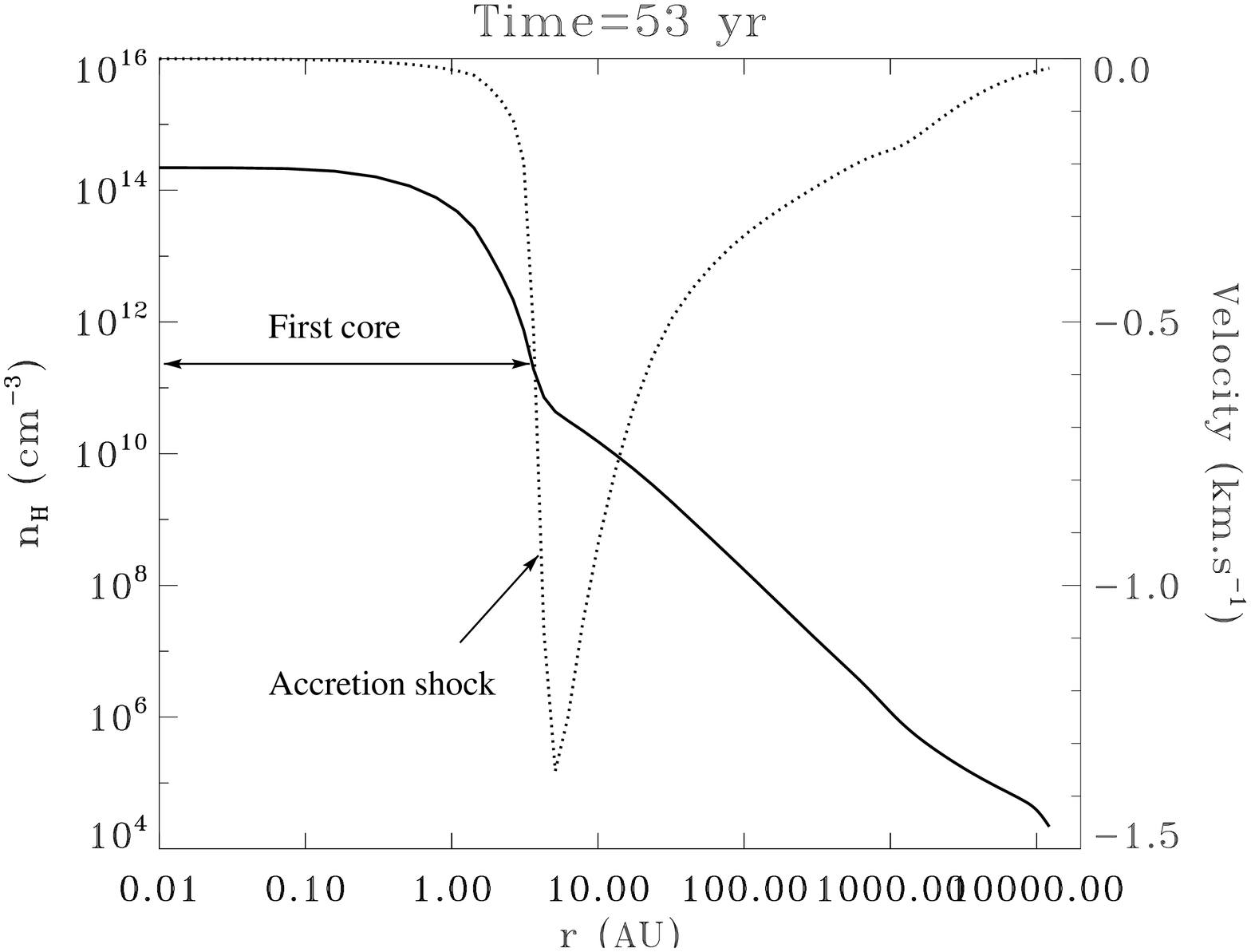,width=8cm,angle=-90}\\
\psfig{file=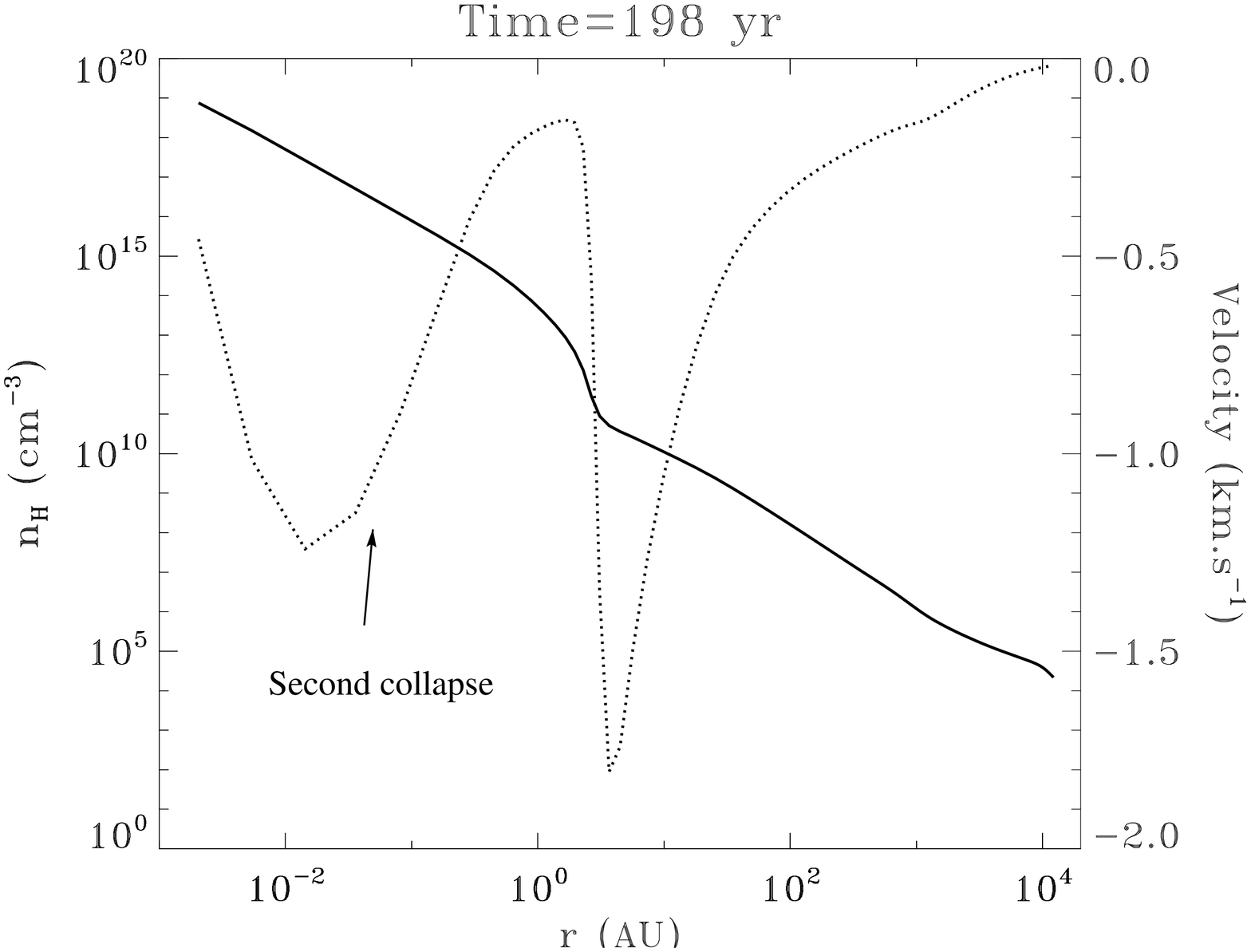,width=8cm,angle=-90}&
\psfig{file=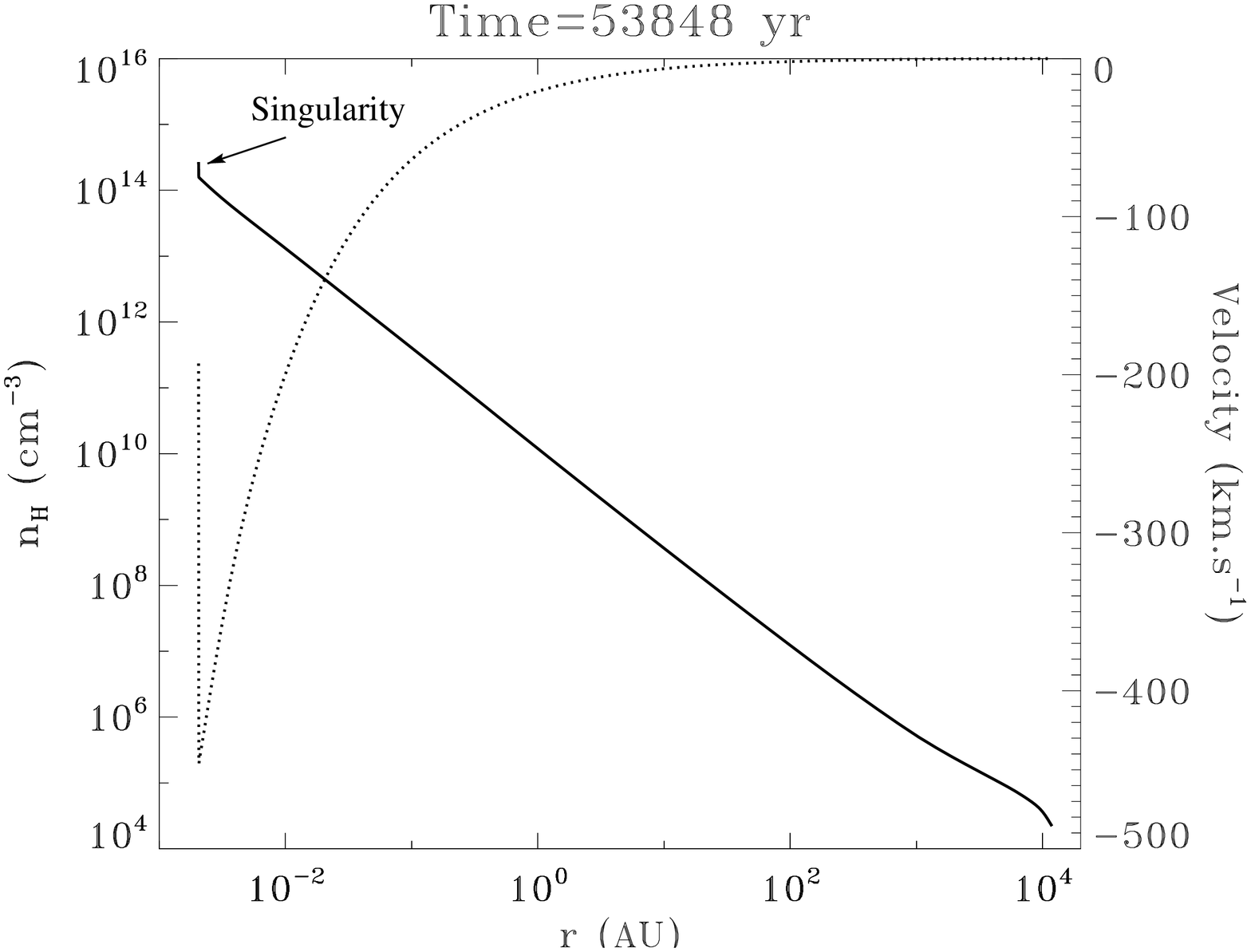,width=8cm,angle=-90}\\
\end{tabular}
\end{center}
 \caption{Six snapshots of the density and velocity profiles
of our reference simulation with cooling. The origin of times is taken as
the formation of the shock. 
}
\label{collapse}
\end{figure*} 

  Figure \ref{collapse} shows the dynamical evolution of the model.
  Features such as the formation of the first adiabatic core and the
  propagation of the expansion wave where the density profile switches
  to a $r^{-1.5}$ law are already described in the literature
  \citep[see][]{L69,W85,M98,MI00}. We concentrate here on the heating
  and cooling processes.

\begin{figure*}[h]
\begin{center}
\begin{tabular}{cc}
\psfig{file=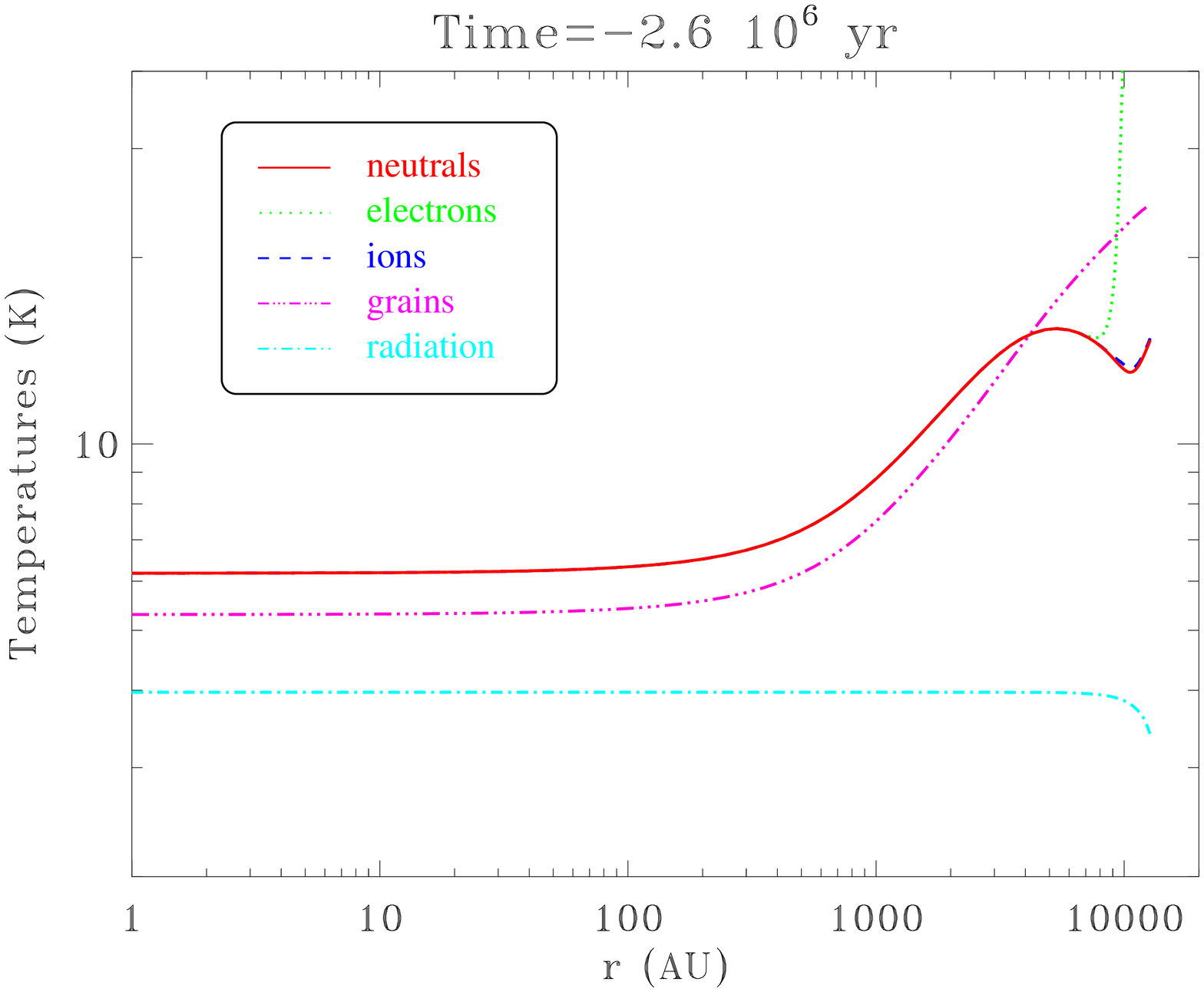,width=8cm,angle=0}&
\psfig{file=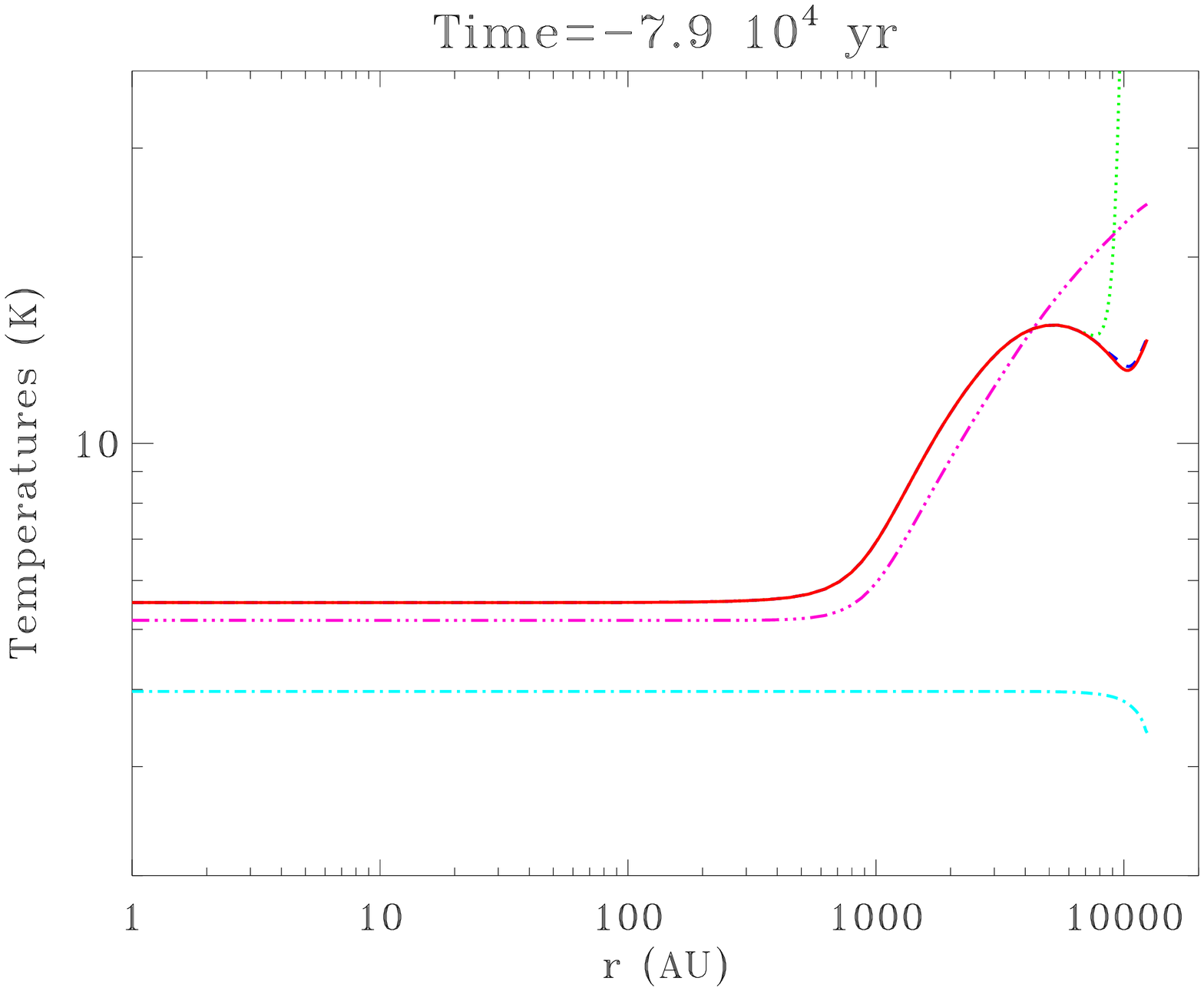,width=8cm,angle=90}\\
\psfig{file=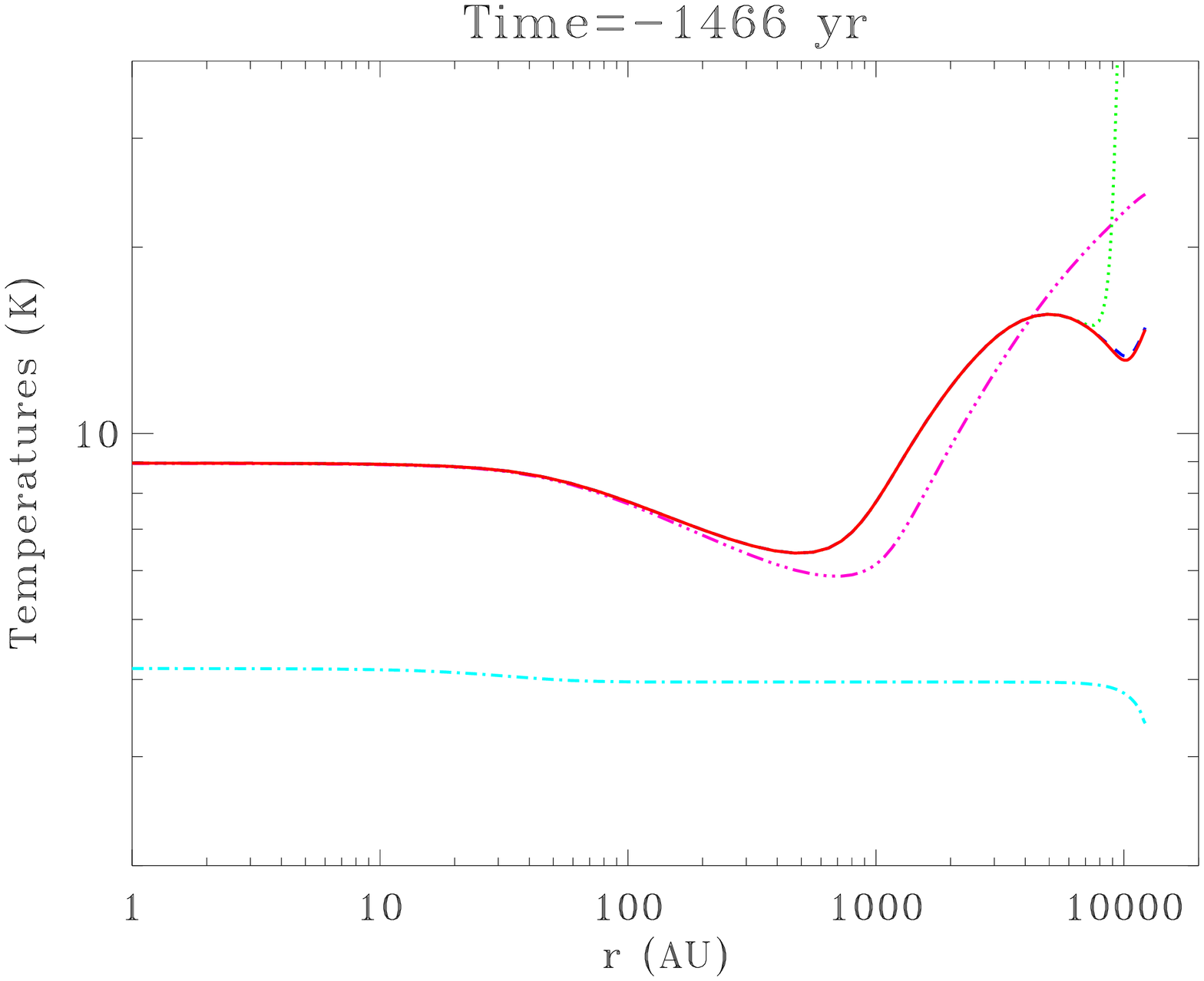,width=8cm,angle=90}&
\psfig{file=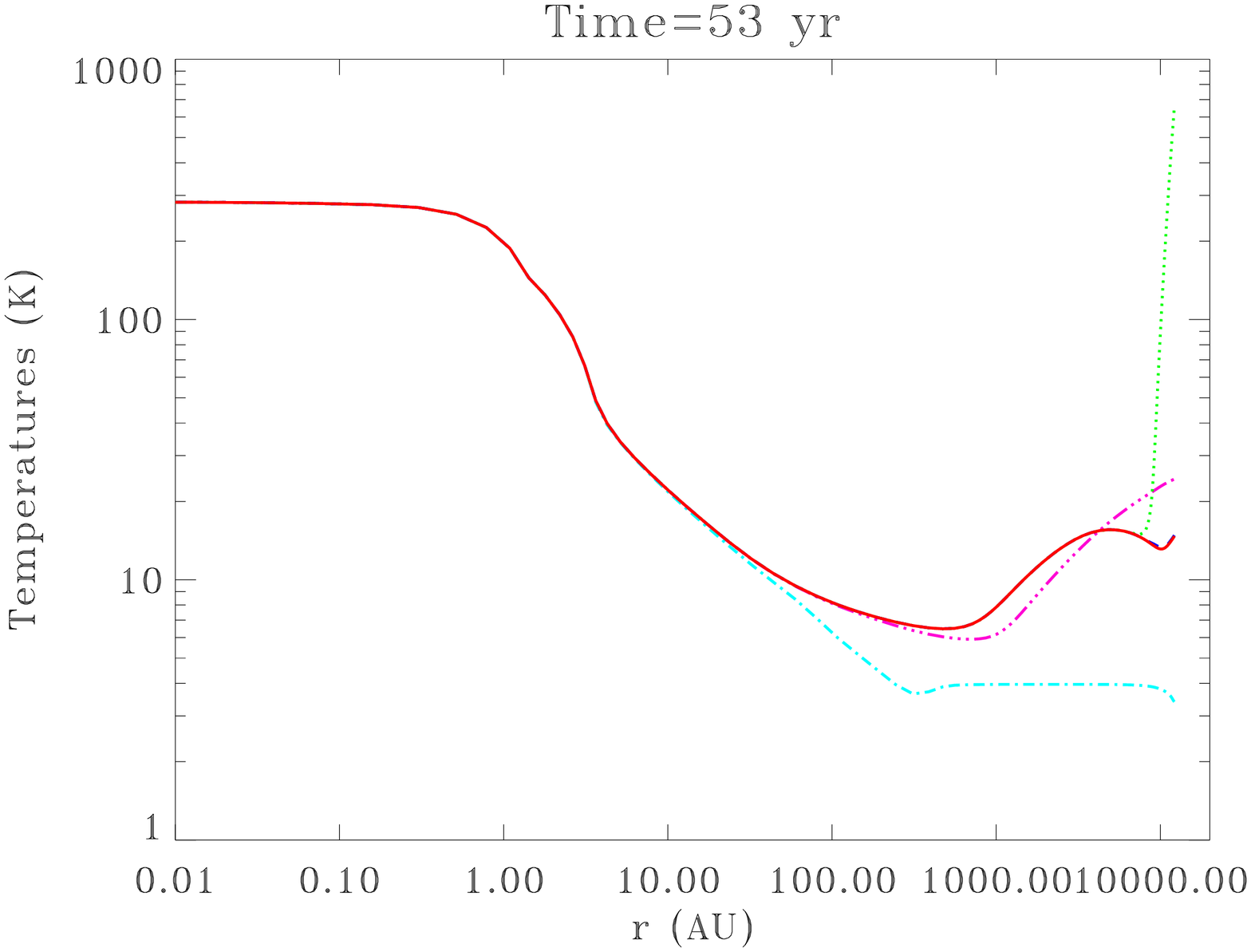,width=8cm,angle=90}\\
\psfig{file=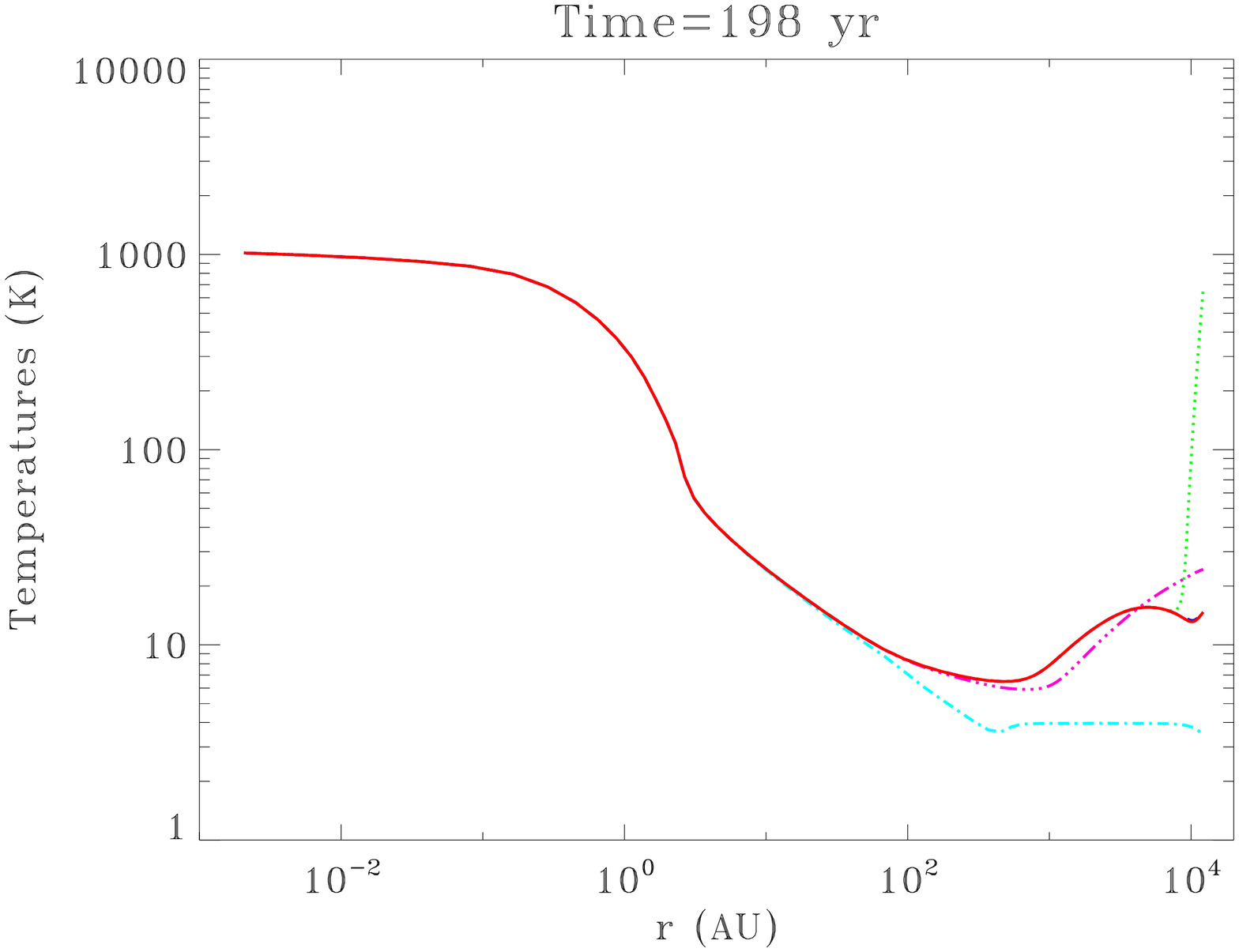,width=8cm,angle=90}&
\psfig{file=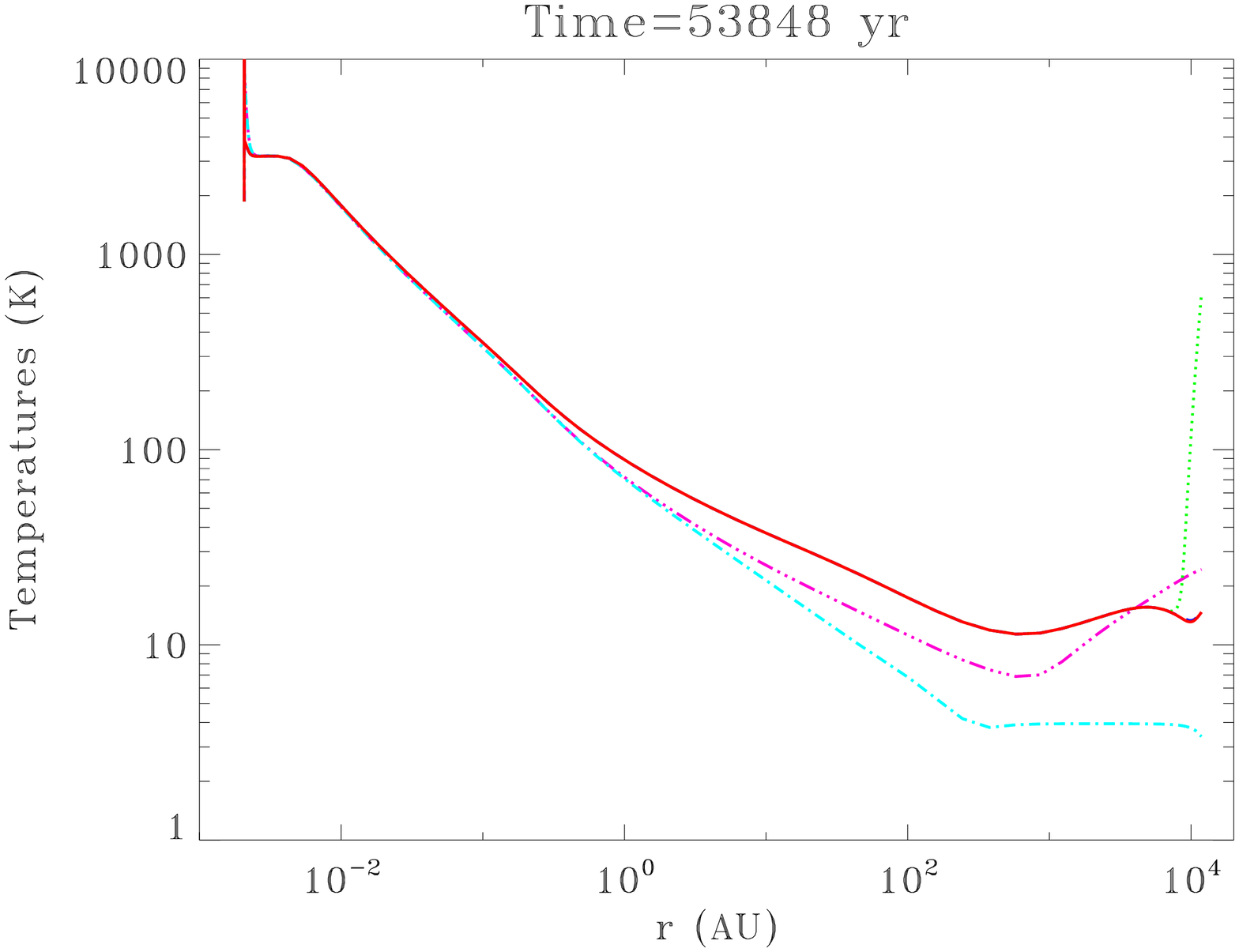,width=8cm,angle=90}\\
\end{tabular}
\end{center}
 \caption{Evolution of the temperature  profiles for the
reference simulation. The times of the six snapshots
 correspond to the six snapshots displayed in figure \ref{collapse} }
\label{temp}
\end{figure*} 

  Figure \ref{temp} shows the evolution of the temperature profiles in
  a sequence of snapshots during the collapse.  The gas temperature at the
  very edge of the envelope is determined by the balance between the
  molecular cooling (largely dominated by CO) and the heating through
  the photoelectric effect. The temperature of the electrons is
  decoupled from the heavier particles as they experience almost all
  the heating. Deeper in the envelope cosmic-ray heating \citep{SS69}
  takes over as the UV-field gets shielded from the outside.

  The grains have no influence on the gas temperature at low
  densities, but the frequency of collisional exchanges between the
  gas and the grains is proportional to the square of the
  density. When the central density rises, these interactions get
  stronger than the collisionally saturated lines and CO ceases to be
  the main cooling agent. The cooling of the gas is then relayed by
  the grains which bring the gas temperature closer to the continuum
  radiation temperature and a dip appears in the temperature
  profile. Deeper in the core and later during the collapse the
  rate of compressional work gets stronger than the cosmic-ray heating and
  the temperature is the result of the balance between the rate of work of the
  pressure and the collisional cooling due to the grains.

  Finally when the grains become optically thick to the continuum
  radiation no radiative process can cool the gas anymore.  Its
  temperature gets higher and the accretion shock takes place. 
We take this time as the origin of time for simulations with
cooling. We estimate it from the instant when the maximum of $p_v/p$
along the profile exceeds a half ($p$ and $p_v$ are the thermal and
viscous pressures).

  Below the accretion shock the adiabatic compression goes on until
  the core temperature is high enough for the collisional dissociation
  of H$_2$ to proceed. The second collapse then occurs. We do not
  describe this phase accurately as our equation of state is for an
  ideal gas and we do not consider the evaporation of the grains.

  The code finally evolves through the singularity of the second
  collapse thanks to our fixed inner boundary. The density decreases
  in places reached by the expansion wave. This lowers the
  coupling of the gas temperature with the radiation temperature
  through the grains as seen in equation (\ref{effective}) and the gas
  temperature goes up.

\subsubsection{Influence of cooling}

  In this section we compare our reference run to the 
isothermal model with same $t_p$=10$^7$~yr.

\begin{figure}[h]
\centerline{
\psfig{file=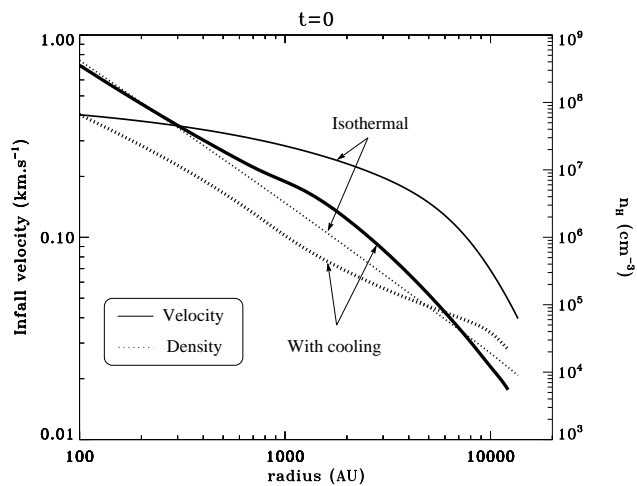,width=9cm,angle=-90}
}
 \caption{Velocity (solid lines) and density (dotted lines) profiles 
at $t=0$ for the reference simulation (thick lines) and an
isothermal simulation with $t_p=10^7$~yr (thin lines). }
	\label{cool}
\end{figure}

  We compare in figure \ref{cool} the velocity and density profiles at
  time $t=0$ in the two simulations: cooling leads to much lower
  velocities and a less steep density profile. The external pressure
  at $t=0$ is larger for the simulation with cooling ($p_{\rm
  ext}(0)=7.6 \times 10^4$~K.k$_{\rm B}.$cm$^{-3}$ versus $p_{\rm
  ext}(0)=4.7 \times 10^4$~K.k$_{\rm B}.$cm$^{-3}$), in agreement with
  the corresponding result on marginally stable states found by
  \cite{G02}. Since we start both simulations with the same initial
  $p_{\rm ext}=3 \times 10^4$~K.k$_{\rm B}.$cm$^{-3}$, the simulation
  with cooling lasts about twice as much time.

\begin{figure}[h]
\centerline{
\psfig{file=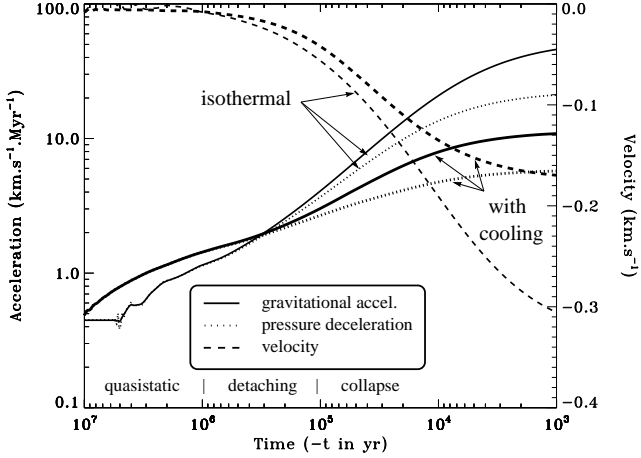,width=9cm,angle=-90}
} 

 \caption{Evolution with time (-$t$) of the gravitational acceleration
 ($Gm/r^2$ solid), the pressure deceleration (-1/$\rho$ d$p$/d$r$,
 dotted) and the velocity (dashed) of the 0.15~M$_\odot$ mass shell for
 the reference run (thick lines) and the $t_p=10^7$~yr isothermal run
 (thin lines). The quasistatic, detaching and collapse phases are
indicated.}
\label{accel}
\end{figure}

 To better understand the difference in velocities we trace the
 Lagrangian particle located at the shell of mass 0.15~M$_\odot$ in
 both simulations. Its velocity evolves in time according to the
 competing gravitational acceleration and pressure deceleration (see
 figure \ref{accel}). The net acceleration $\ddot{r}$ provides us with
 a typical timescale $t_r=\sqrt{-r/\ddot{r}}$ for the change in radius
 which turns out to be of the same order as or lower than the velocity time scale
 $t_u=\dot{r}/\ddot{r}$. We consider also $t_p$ and $t_d$, the sound crossing
 time of the cloud (roughly
 $t_d=10^4$~AU/0.2~km~s$^{-1}=2.4 \times 10^5$~yr).  We distinguish three
  phases:

\begin{enumerate}
 \item When the pressure increase is switched on, low amplitude
 oscillations of period a few $t_d$ characterise the adjustment of
 the cloud. When they finally damp, the hydrostatic equilibrium is
 verified at all times: $t_r \gtrsim t_p > t_d$.  Until about $t=-10^6$~yr, the
 radius and velocity of the Lagrangian particle are hence determined
 by the succession of hydrostatic equilibria driven by the time
 evolution of $p_{\rm ext}$: we refer to this phase as the quasistatic
 phase.
 
 \item For times $-10^6$~yr$ < t < -10^5$~yr, $t_p > t_r > t_d$.  The
 time scale for the variation of the radius is still greater than the sound
 crossing time and the pressure has time to adjust. The
 hydrostatic equilibrium is hence still rather well verified, but $p_{\rm
 ext}$ is no longer controlling the time scales. $p_{\rm ext}$ has
 forced the hydrostatic equilibrium past the marginally stable state,
 and the cloud is slowly starting to collapse.  We refer to this phase
 as the detaching phase.

 \item For $t>-1~10^5$~yr, $t_r$ gets below $t_d$. The velocity and
 radius changes are too fast for the cloud to adjust its pressure
 gradients.  An increase of the gravitational acceleration reflects in
 a change of radius which in turn yields a higher acceleration. This
 runaway is hence sensitive to its initial conditions in radius and
 velocity, determined at the end of the detaching phase. We refer to
 this last phase as the collapse phase.
\end{enumerate}

  We note that the transition times remain within 50\% of these estimates
for mass shells 0.3 and 0.6~M$_\odot$ in the case with cooling
(with higher masses collapsing later). But the transition times can be
twice earlier in the isothermal run compared to the run with cooling
at shell mass 0.6~M$_\odot$.

\begin{figure}[h]
\centerline{
\psfig{file=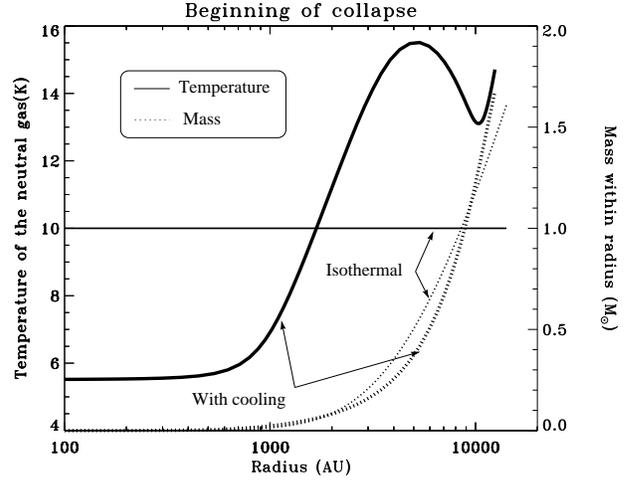,width=9cm,angle=-90}
} \caption{Temperature of the neutral gas (solid lines) and mass
within radius (dotted lines) at the beginning of the collapse (for a
given density contrast of 100, around time $t=-9~10^4$~yr) for the
reference run (thick lines) and the isothermal simulation with
$t_p=10^7$~yr (thin lines).}
\label{stat}
\end{figure}

 We show a comparison of the isothermal simulation and the reference
 run in figure \ref{stat} at the very beginning of the collapse, ie:
 the last hydrostatic configuration. The collisional exchanges between
 gas and grains are responsible for a dip in the temperature profile in
 the inner 0.2~M$_\odot$.  Pressure gradients depend on the density
 and temperature gradients:
\begin{equation}
  \frac1{\rho}\dpart{p}{r}=\frac{k_{\rm B}}{\mu}(T\dpart{\ln \rho}{r}
+\dpart{T}{r})
\end{equation}
 In the isothermal case, only the density term remains (first
 term). In the case with cooling, the temperature gradients (second
 term) are also at play.  Due to the temperature dip, the temperature
 gradient term is maximum at the mass shell 0.05 M$_\odot$, where it
 is of the same order as the isothermal pressure gradient term, but
 directed outward. However, due to a steeper density profile in the
 mass range $[0,0.05]$~M$_\odot$, the density term is roughly twice
 the isothermal pressure gradient. In effect, the resulting pressure
 gradient is actually of the same order as in the isothermal case
 \citep[as noted by][ for the marginally stable state]{G02}. It is
 nevertheless slightly greater in the innermost 0.05 M$_\odot$ and the
 outermost 0.6 M$_\odot$, and slightly lower (by about 35\% at most)
 in the rest of the mass.  As a result, the bulk of the envelope (0.05
 M$_\odot < m < 1.1~$M$_\odot$) is more extended (see figure
 \ref{stat}).

  This slight difference is increased during the collapse.
Since these Lagrangian particles in the isothermal run start the
collapse phase at a lower radius, they experience a higher
gravitational acceleration, hence they reach sooner lower radii and
their gravitational acceleration becomes even larger compared to the
same Lagrangian particle in the reference run.

As a result the protostellar infall velocities at a given mass are
much larger in the isothermal case than in the case with cooling.
This explains the much lower supersonic mass fraction for the
reference run compared to isothermal cases as seen in figure
\ref{x2}. However, to compare velocities at a given radius, one needs to keep
in mind that a given mass is found at a smaller radius in the
isothermal case. But this only partly compensates for the difference
on the Lagrangian velocities and the net result is that infall velocities
 are larger at a given radius and a sufficiently late time for
the isothermal run.

Finally, we note that the reasoning ``the gas cools down, hence there
is less support, hence the collapse is more violent'' appears to be
wrong. First, the quasistatic sequence of equilibria approaching the
marginal state is modified in a non trivial way by the cooling: a
given external pressure increase yields less contraction when cooling
is switched on. As a result, the isothermal contraction motions are
higher in the quasistatic and detaching phases. Second, the
temperature dip in the last (unstable) hydrostatic equilibrium leads
to a more extended inner envelope.  Third, the collapse is sensitive
to the configuration (velocities and radius) in the last hydrostatic
state: it is much milder in the case with cooling.

\subsubsection{Influence of other parameters}
\label{parameters}

  We varied $t_p$ between 10$^6$ and 10$^8$ yr without noticing a
  significant change (the supersonic mass fraction varies at most by
  20\% in relative value). This is in contrast to the isothermal
  models and suggests that indeed the thermal structure is
  constraining the collapse.

  We reckon that an important factor is the depth and steepness of the
  temperature dip because this regulates the pressure gradients in the
  marginally stable state. This is partly determined by the minimum
  temperature in the dip, which in turn depends on the external
  radiation field. We tested $ G_0 =0.5$ which slightly lowers the
  temperature dip and yields a small decrease (of about 10\% ) in the
  supersonic mass fraction. We also investigated much bigger changes
  in the value of $T_r$ from 2.7 to 15~K while keeping $ G_0 $ set to
  zero.  Indeed, the minimum grain temperature is always about 1~K
  above $T_r$ when $ G_0 =0$. This provides an easy way to control the
  temperature profile, even if the values we use do not correspond to
  practical situations (remember that $T_r$ is the {\it bolometric}
  temperature of the ambient infrared radiation field). Results show
  that velocities comparable to the infall motions in isothermal
  models are recovered for high values of $T_r$ (above 10~K) which
  flatten the dip in temperature.

  From section \ref{reference} we note that CO is the only molecule
  that gives rise to significant cooling in the envelope. In addition,
  the infall motions are so slow and the densities so high that the
  chemical steady state is obtained everywhere at all times for
  CO. Indeed, the adsorption time for the CO molecule is given by
\begin{equation}
t_{\rm ad}=\Sigma_d n_H \sqrt{\frac{8 k_{\rm B} T_g}{\pi m_{\rm CO}}}
\end{equation}
where $\Sigma_d$ is the effective surface of grains per H nucleus and
$m_{\rm CO}$ is the weight of the molecule CO.
 On the other hand, the time scale for collapse can be estimated from
the local free-fall time scale
\begin{equation}
t_{\rm ff}=\sqrt{\frac{3 \pi}{16 G \mu n_H}}
\end{equation}
where $G$ is the gravitational constant.  The ratio $t_{\rm ff}/t_{\rm
ad}$ is greater than 1 for 
\begin{equation}
n_H > 9.4 \times 10^4 {\rm cm}^{-3}.{\rm K}/ T_g
\end{equation}
 Since the densities at the edge of the cloud are greater than
 $10^4$~cm$^{-3}$ in the marginally stable state, the time scales for
 adsorption are significantly shorter than the compression time
 scales.  This allows to compute the CO abundance from the equilibrium
 between adsorption and desorption onto and from grains with virtually
 no loss of accuracy for the cooling function of the gas. Since all C
 in gas phase is usually locked in CO, we conclude that
 out-of-equilibrium chemistry has no impact on the cooling in the
 envelope.

 Finally, we ran a simulation with a total mass of 3.4~M$_\odot$
 (double of the reference run, every other parameters being
 fixed). The density profiles are very similar at all times, the more
 massive being only an extension of the reference run at larger
 radii. As a result, the temperature dips at the beginning of the
 collapse are also very similar for radii smaller than 4000~AU (or
 0.2~M$_\odot$).  At the formation of the stellar object, the
 innermost infall motions for the massive run are only about 20\%
 higher than the reference model. The outermost infall motions are
 much higher (by a factor greater than 2 outside the shell of mass
 0.2~M$_\odot$). The scaling relations expected for isothermal models
 are hence not valid for the velocity profiles.

\section{Comparison with observations}
\label{cobs}
\subsection{IRAM~04191}
 
IRAM~04191+1522 -- hereafter IRAM~04191 for short -- is one of the
youngest low mass class 0 protostars known so far in the Taurus
molecular cloud ($d = 140$ pc).  It features a prominent envelope
($\sim 1.5$ M$_\odot$), a powerful bipolar outflow and a very
low bolometric luminosity of $L_{\mbox{\tiny bol}} \sim 0.15$
L$_{\odot}$ \citep{A99}. \cite{A99} estimated an age of 1-3
$\times 10^4$ yr since the beginning of the accretion phase.
\citet{B02} performed a detailed analysis of the molecular line emission 
observed with the IRAM 30m telescope and derived strong constraints on
the velocity structure of the envelope (see, e.g., the shaded area in
Fig.~\ref{compIRAM}). They showed that the envelope is undergoing both
extended infall motions and fast, differential rotation. They proposed
that the inner envelope ($r < 3000-3500$ AU) corresponds to a
magnetically supercritical core decoupling from an environment still
supported by magnetic fields.

Being born in the Taurus molecular cloud where star formation seems to
proceed in a rather isolated and quiet manner, IRAM~04191 is likely to
have experienced a slowly varying background in the past. It is
therefore appropriate to compare it with models of collapse triggered
quasistatically.

\subsection{Isothermal models}

\label{compiso}
\begin{figure}[h]
\centerline{
\psfig{file=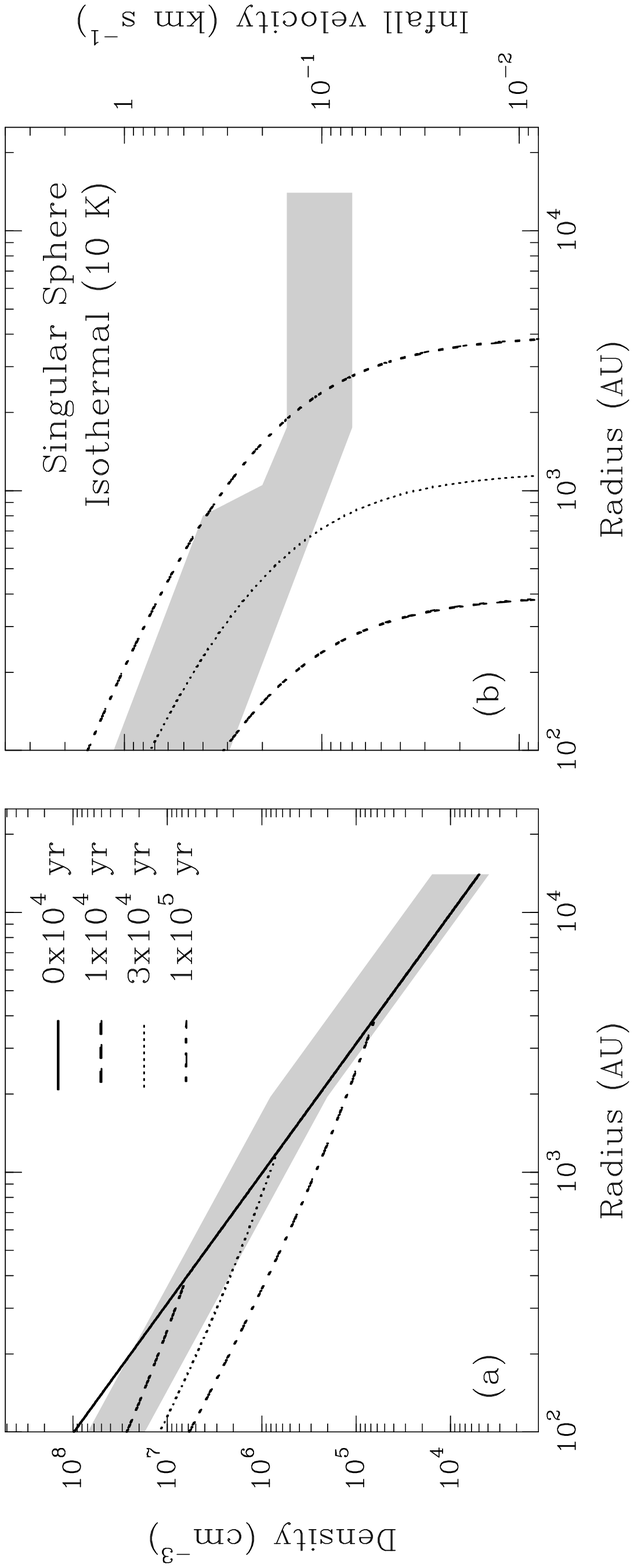,width=9cm,angle=-90}
}
\centerline{
\psfig{file=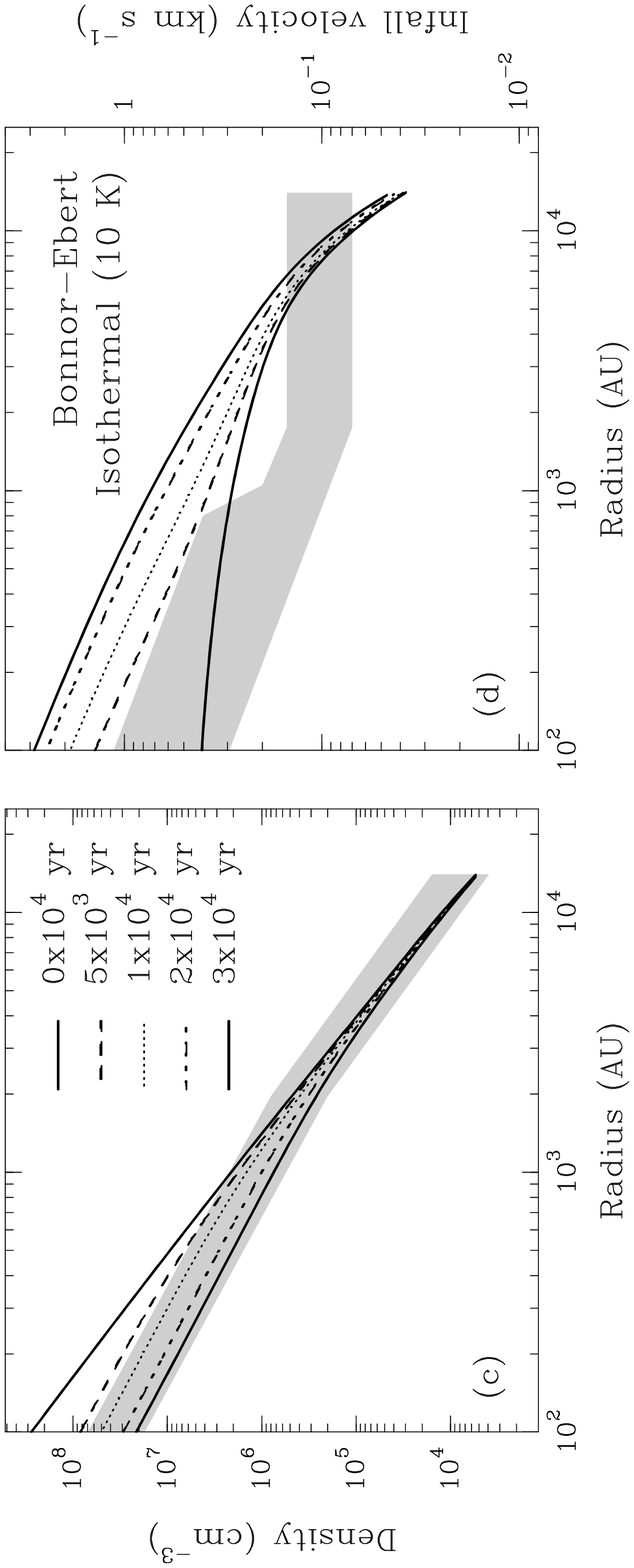,width=9cm,angle=-90}
} \caption{Comparison of isothermal models to the observational
constraints derived for IRAM~04191 by \citet{B02}. {\it Upper panel (a,b):}
initial conditions are a singular isothermal sphere \citep{S77}. {\it
Lower panel (c,d):} initial conditions are a critical Bonnor-Ebert sphere.
}
\label{compIRAM}
\end{figure}

Figure \ref{compIRAM} plots the results of two isothermal computations
against the observational constraints of \citet{B02} for the density
and velocity profiles. The two simulations differ only by their initial
conditions.  

The first simulation (figure \ref{compIRAM}a and b) is the self-similar solution
obtained by \citet{S77} with a singular isothermal sphere for initial
conditions. The density profiles can fit the observational
constraints for $t=0$ to 3~10$^4$~yr but it shows no infall
velocity in the outer part of the envelope for these ages.

The second simulation (figure \ref{compIRAM}c and d) is the least
dynamical collapse we could achieve for a Bonnor-Ebert sphere
($t_p=10^9$~yr).  The density profile agrees quite well with the
observational constraints at the estimated age of IRAM~04191 (from
10$^4$ to $3 \times 10^4$~yr).  But the velocities are much too high
around a radius of 3000~AU at $t=0$ and this is even worse for the
later relevant ages. The isothermal collapse is therefore way too
dynamical to account for the observed infall velocities.  We plot the
results for a temperature $T=10$~K and a total mass $M=1.7$~M$_\odot$,
but the rescaling relations show that varying these parameters within
their range of uncertainty \citep[$T$ between 7 and 15~K, $M$ between
1 and 2 M$_\odot$ according to][]{A99} has minor effects on the
profiles.
  
An even more stringent constraint from the observations by \citet{B02}
is~: 10\% at most of the mass of IRAM~04191 has supersonic motions.
However, even for the least dynamical of our isothermal simulations
the supersonic mass fraction at time $t=0$ is still 31\% and much more
at the estimated age of IRAM~04191 (see figure \ref{x2}).

\subsection{Models with cooling and chemistry}

\begin{figure}
\centerline{
\begin{minipage}[c]{5cm}
\psfig{file=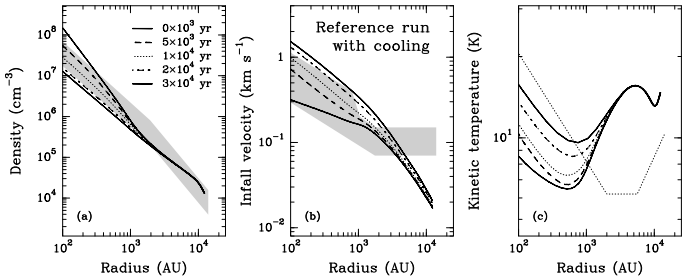,width=9cm,angle=-90}
\end{minipage}
} 

\caption{ Comparison of the reference model with \citet{B02}'s
observational constraints on IRAM~04191.  }

\label{compobs}
\end{figure}

  When we include the cooling, the slower infall motions agree much
  better with the observational constraints inside a radius of 3000~AU
  as seen in figure \ref{compobs}. In particular, the supersonic mass
  fraction is now much lower and in much better agreement with the
  observations (see figure \ref{x2}).  However, there still remains
  a big discrepancy outside this radius, where the observations
  require an almost constant infall velocity profile.
 
\begin{figure}
\centerline{
\begin{minipage}[c]{5cm}
\psfig{file=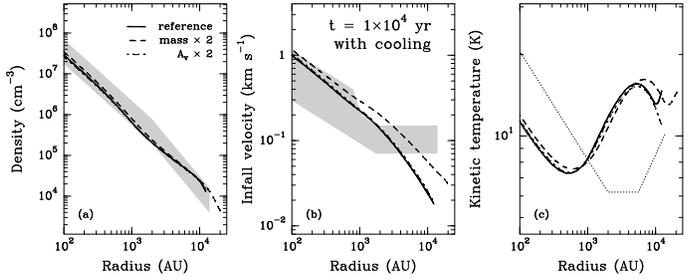,width=9cm,angle=-90}
\end{minipage}
} 

\caption{ Comparison of models with cooling at $t=10^4$~yr with \citet{B02}'s
observational constraints on IRAM~04191.}

\label{compar}
\end{figure}

  Simulations with cooling also yield a kinetic temperature of the
  gas closer to the profile deduced by \cite{B02} from the analysis of
  molecular line spectra: in summary, they need a gas temperature of 6-7~K
  in the range of radii 2000-6000~AU and a gas temperature at the edge of
  around 10~K. In this respect, our minimum temperature (around 7~K)
  is too close from the centre of the core and our temperature at the
  edge of the core is slightly too hot (15~K). A higher $A_v$ at the surface
  only partly remedies to the latter problem (see figure
  \ref{compar}): a bump in temperature at around 6000~AU still
  remains. This is due to the cosmic-ray heating. In the present work
  we use an old formulation due to \cite{SS69} and we assume this
  process heats the neutral gas directly.  However, the realistic
  mechanism rather involves heating of the electron gas which in turn
  exchanges heat via collision with neutrals. This could help decrease
  the temperature of the neutral gas if the coupling to electrons
  remains low. Future refinements of the code will include a proper
  treatment of the electron exchanges between the gas and the grains.

  The density profiles are not very sensitive to the model of collapse 
and they remain within the observational constraints. Density
and velocity constraints on our models yield an independant estimate
of the age of IRAM~04191 in the interval 0.5-2~10$^4$~yr.

\section{Discussion}
\subsection{The temperature dip}
\citet{MI00,E01,Z01,G02} also find a dip in the temperature profile as a result
of their model for the grains energy budget. However, all these
authors neglect the collisional exchanges between gas and grains to
compute $T_d$.  As pointed out in section \ref{coll}, this leads to a
coupling of $T_d$ and $T_g$ as soon as the density is above
10$^5$~cm$^{-3}$.  Besides, $T_d$ is then tied to the local radiation
field, which is shielded near the edge of the cloud. As a result, the
temperature starts to decrease for much larger radii than in our
model. The temperature gradients are hence less steep and spread out
to larger radii. Our model should then also account better for the flat
temperature profiles found outward of 1000~AU for the prestellar cores
L1498 and L1517B \citep{T04}.

Grain opacities and gas-grain coupling are hence the main physical
ingredients that control the position of the temperature dip. Another
important factor is the cosmic-ray heating, as discussed in section
\ref{cobs}.

\subsection{Magnetic fields and rotation}
 \citet{B02} observe two regimes of infall and rotation in the outer
 envelope and inside a radius of 3000~AU. They proposed that the outer
 part of the object is still subcritical and supported by the magnetic
 field, but the collapsing inner part is supercritical.  Although we
 did not include any magnetic field, we suggest that a slow increase
 of the external pressure can mimic the slow diffusion of the magnetic
 field from the outer parts. Hence our study applies to the inner core
 which frees itself from the magnetic field. A more consistent
 treatment including a magnetic field and differential rotation is
 desirable but hard to implement in practice due to the loss of
 spherical symmetry.

\subsection{Turbulent support}

The observed macroscopic random motions amount to a significant
fraction of the broadening of the lines \citep{B02}. However,
the origin of these turbulent motions is still not well understood and
their role in the collapse is even less well known.

\subsection{Pressure variation time scale}
 It is not clear whether the picture of quasistatically collapsing
clouds matches the observations of the turbulent interstellar medium
that surrounds them. Our results hold for pressure time scales down to
$t_p$=1~Myr.  However, it is a rather difficult task to estimate the
variability of the turbulent background in which the cloud is
embedded.  

Larson relations for a scale of 0.2~pc (corresponding to
 IRAM~04191's diameter) give a velocity dispersion of roughly
0.5~km.s$^{-1}$. This yields a time variability of 0.4 Myr. 

C$^{18}$O(1-0) observations of \citet{B02} with the IRAM 30m
telescope, which probe the low density outskirts of IRAM~04191
($n\simeq 3\times 10^3~$cm$^{-3}$), show a turbulent velocity
dispersion of 0.25 km.s$^{-1}$ (FWHM 0.60 km.s$^{-1}$). We extrapolated this
velocity dispersion from the beam diameter (3500~AU) to 0.2 pc with a
Kolmogorov scaling (velocity $\propto$ length$^\frac13$) and found a
time scale of the order of 0.35~Myr.

These time scales are hence less than half the lowest
value we tried for $t_p$. However, we note that they are valid for the
{\it intra}-cloud velocity dispersion {\it nowadays}. They hence
represents only a lower bound for the time-scale of the {\it
inter}-cloud medium during the early collapse phase.

\section{Summary and conclusions}

We investigated the collapse of spherical clouds driven by a slow
increase in external pressure. We compared isothermal models
to simulations including cooling and found that the sequence of
hydrostatic equilibria controls the infall motion in both the
protostellar and the prestellar phases. In particular, the last
(unstable) hydrostatic equilibrium reached controls the infall
velocities during the collapse.

First, we note that hydrostatic equilibrium holds past the
marginal state up to density contrasts of  a factor of 10 greater
than in the marginally stable state in our reference simulation. 
A collapsing cloud still in the detaching phase can thus exhibit an
{\it unstable} hydrostatic density profile. Unstable hydrostatic
equilibria should then be good models for observations of slowly
contracting prestellar cores.  Second, at the beginning of the
collapse, the coupling between the gas and the background radiation
mediated by the grains yields a temperature dip in the inner envelope
which slightly lowers the pressure gradients. As a result, the last
hydrostatic configuration is more extended.  Third, the collapse phase
is very sensitive to the initial conditions, hence to the last
hydrostatic state. As a result, the collapse is much milder than in
the isothermal case. We therefore suggest that the external radiation
field and the collisional coupling between gas and grains are key
ingredients constraining the dynamics of the collapse.

Our computations include a full out-of-equilibrium chemical network,
but we show that only the adsorption and desorption of CO on grains
have an impact on the cooling function. These processes
can be safely considered at equilibrium for the range of parameters 
investigated. The energy budget is thus dominated by radiative
transfer and microphysics on grains.

We compared the results of our computations to observational
constraints obtained by \cite{B02} for the class 0 protostar
IRAM~04191. We show that even the most quasistatic isothermal collapse
is unable to account for the slow infall motion observed.  On the
other hand, simulations with cooling not only improve the kinetic 
temperature but also provide better velocity profiles in the inner
3000~AU. Rotation and magnetic fields are likely to be at play in the
outer regions, but are hard to account for in 1D simulations. Future
investigations should aim at describing ambipolar diffusion in a
non-isothermal context.

This study justifies the use of post-processed chemistry
using a prescribed dynamical background, keeping in mind that the line
profiles directly probe infall motions and excitation temperatures.

\begin{acknowledgements}
We are indebted to Prof. G. Pineau des Forêts for suggesting to us the
model for grain adsorption and desorption. We thank the anonymous referee
for a constructive report and interesting comments.
\end{acknowledgements}

\bibliographystyle{aa}

\end{document}